\documentclass[a4paper]{article}
\usepackage[hmarginratio=1:1,top=20mm,columnsep=20pt]{geometry} 
\usepackage{graphicx}
\usepackage{abstract} 
\usepackage{titling} 
\usepackage{titlesec} 
\def\anti_particle{\tilde}
\def\cls{CL$_s$ }

\date{}

\pretitle{\begin{center}\Huge\bfseries} 
\posttitle{\end{center}} 
\title{Recent results of the DANSS experiment}
\author{M.~Danilov on behalf of the DANSS Collaboration\\
\normalsize Lebedev Physical Institute of the Russian Academy of Sciences,\\ 
\normalsize 53 Leninskiy Prospekt, Moscow, 119991, Russia\\ 
\normalsize {danilov@lebedev.ru}\\
}

\renewcommand{\maketitlehookd}{%
\begin{abstract}
DANSS is a highly segmented 1~m${}^3$ plastic scintillator detector. Its 2500 one meter long scintillator strips have a Gd-loaded reflective cover. The DANSS detector is placed under an industrial 3.1~$\mathrm{GW_{th}}$ reactor of the Kalinin Nuclear Power Plant 350~km NW from Moscow. The distance to the core is varied on-line from 10.7~m to 12.7~m. 
Recent results on searches for a sterile neutrino are presented as well as measurements of the antneutrino spectrum dependence on the fuel composition. All results are preliminary.\\

{\noindent
Talk presented at the La Thuile 2018 Conference.\\
PACS: 14.60.Pq, 14.60.St
}
\end{abstract}
}

\begin{document}

\maketitle

\section{Introduction}

The number of active neutrinos is limited to 3 by
 the measurements of the Z boson decay width \cite{PDG}. However, existence of additional sterile neutrinos is not excluded. Several effects observed with about $3\sigma$ significance level can be explained by active-sterile neutrino oscillations. The GALEX and SAGE Gallium experiments performed calibrations with radioactive sources and reported the ratio of numbers of observed to predicted events of $0.88\pm 0.05$ \cite{SAGE}. This deficit is the so called ``Gallium anomaly'' (GA). Mueller et al. \cite{Mueller} made new estimates of the reactor $\anti_particle\nu_e$ flux which are about 6$\%$ higher than experimental measurements at small distances. This deficit is the so called ``Reactor antineutrino anomaly'' (RAA). Both anomalies can be explained by active-sterile neutrino oscillations at very short distances requiring a mass-squared difference of the order of 1~eV$^2$ \cite{Mention2011}. 

The survival probability of a reactor $\anti_particle\nu_e$ at short distances in the 4$\nu$ mixing scenario (3 active and 1 sterile neutrino) is described by a familiar expression

\begin{equation}
1-\sin^22\theta_{14}\sin^2\left(\frac {1.27\Delta m_{14}^2 [\mathrm{eV}^2] L[\mathrm m]}{E_\nu [\mathrm{MeV}]}\right).
\end{equation}

The existence of sterile neutrinos would manifest itself in distortions of the $\anti_particle \nu_e$ energy spectrum at short distances. At longer distances these distortions are smeared out and only the rate is reduced by a factor of $1-\sin^2(2\theta_{14})/2$. Measurements at only one distance from a reactor core are not sufficient since the theoretical description of the $\anti_particle \nu_e$ energy distribution is considered not to be reliable enough. The most reliable way to observe such distortions is to measure the $\anti_particle \nu_e$ spectrum with the same detector at different distances. In this case, the shape and normalization  of the $\anti_particle \nu_e$ spectrum as well as the detector efficiency are canceled out. Detector positions should be changed frequently enough in order to cancel out time variations of the detector and reactor parameters. The DANSS experiment uses this strategy and measures $\anti_particle \nu_e$ spectra at 3 distances from the reactor core centre: 10.7~m, 11.7~m, and 12.7~m to the detector centre. 
The detector positions are changed typically 3 times a week. Antineutrinos are detected by means of the Inverse Beta Decay (IBD) reaction
\begin{equation}
\label{eq1}
\anti_particle{\nu}_e + p \rightarrow e^+ + n ~\mbox{with}~ E_{\anti_particle\nu} = E_{e^+} + 1.80~\mathrm{MeV}.
\end{equation}

\section{The DANSS Detector}
The DANSS detector was constructed by the ITEP-JINR collaboration. It is described elsewhere \cite{DANSS}. Here we mention only a few essential features. DANSS is installed under the core of a 3.1~GW$_{\rm th}$  industrial power reactor at the Kalinin Nuclear Power Plant (KNPP) 350 km NW of Moscow. The reactor materials provide a good shielding equivalent to $\sim$~50~m of water, which removes the hadronic component of the cosmic background and reduces the cosmic muon flux by a factor of 6. The size of the reactor core is quite big (3.7 m in height and 3.2 m in diameter) which leads to the smearing of the oscillation pattern. This drawback is compensated by a high $\anti_particle{\nu_e}$  flux. 

DANSS consists of 2500 polystyrene-based extruded scintillator strip ($1 \times 4 \times 100$~cm${}^3$) with a thin ($\sim 0.2$~mm) Gd-containing surface coating. The amount of Gd in the detector is 0.35$\%_{wt}$. The coating serves as a light reflector and a ($n,\gamma$)-converter simultaneously. The strips are arranged in 100 layers of 25 strips. The strips in the adjacent layers are orthogonal. The detector is placed inside a composite shielding of copper (5~cm), borated polyethylene (8~cm), lead (5~cm) and one more layer of borated polyethylene (8~cm). It is surrounded on 5 sides (excluding bottom) by double layers of 3~cm thick scintillator plates to veto cosmic muons.

Light from the strip is collected with three wavelength-shifting (WLS) Kuraray fibers Y-11, $\oslash$~1.2~mm, glued into grooves along the strip. 
The central fiber is read out with a Silicon PhotMultiplier (SiPM) (MPPC S12825-050C(X)).  The side fibers from 50 parallel strips (a module – 10 layers with 5 strips each) are read out with a compact photomultiplier tube (PMT) (Hamamatsu R7600U-300).   So that the whole detector (2500 strips) is a structure of 50 intercrossing modules. 
All signals are digitized with specially designed 12~bit, 125~MHz FADCs\cite{FADC}. Only PMTs, SiPMs, and front-end electronics are placed inside the shielding but outside the Cu layer.
 SiPMs (PMTs) register about 18 (20) photo-electrons (p.e.) per MeV. These numbers were obtained using measurements with cosmic muons and artificially driven LEDs. So the total number is 38 p.e./MeV. 
Parameterized strip response non-uniformities have been incorporated into the GEANT4 (Version 4.10.4) MC simulation of the detector. The MC simulation included also a spread in the light yields of different strips, dead channels, Poisson fluctuations in the number of p.e. at the first 2 PMT dinodes, the excess noise factor for SiPMs due to the optical cross-talk between pixels. The experimental energy resolution for cosmic muon signals in the scintillator strips is 15$\%$ worse than that from the MC calculation.
Therefore, the MC estimations are scaled up by the corresponding factor.
After that, the MC describes well the detector response for different radioactive sources. 

\begin{figure}[th]
\centering
\includegraphics[width=0.45\linewidth]{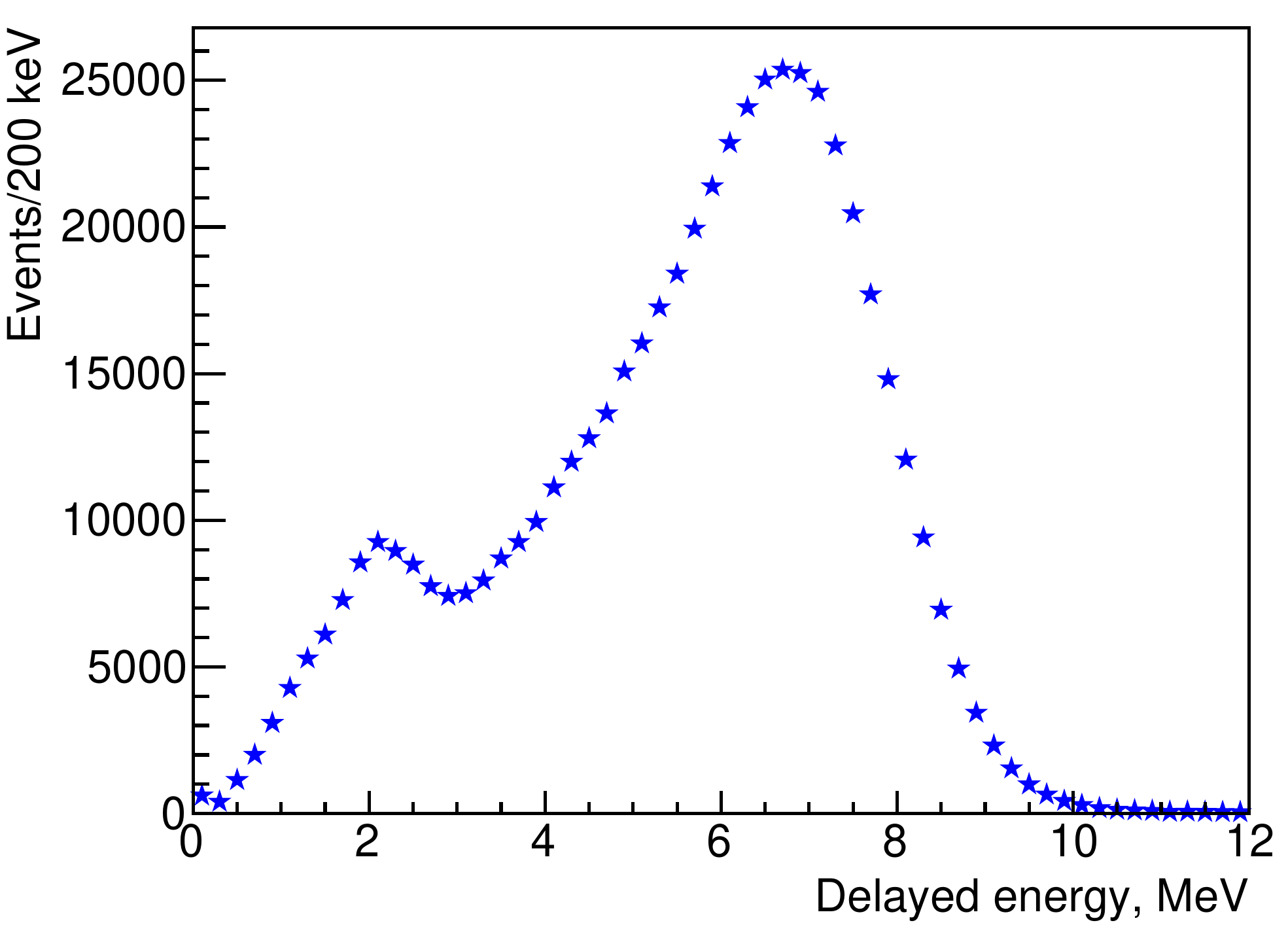}
\includegraphics[width=0.45\linewidth]{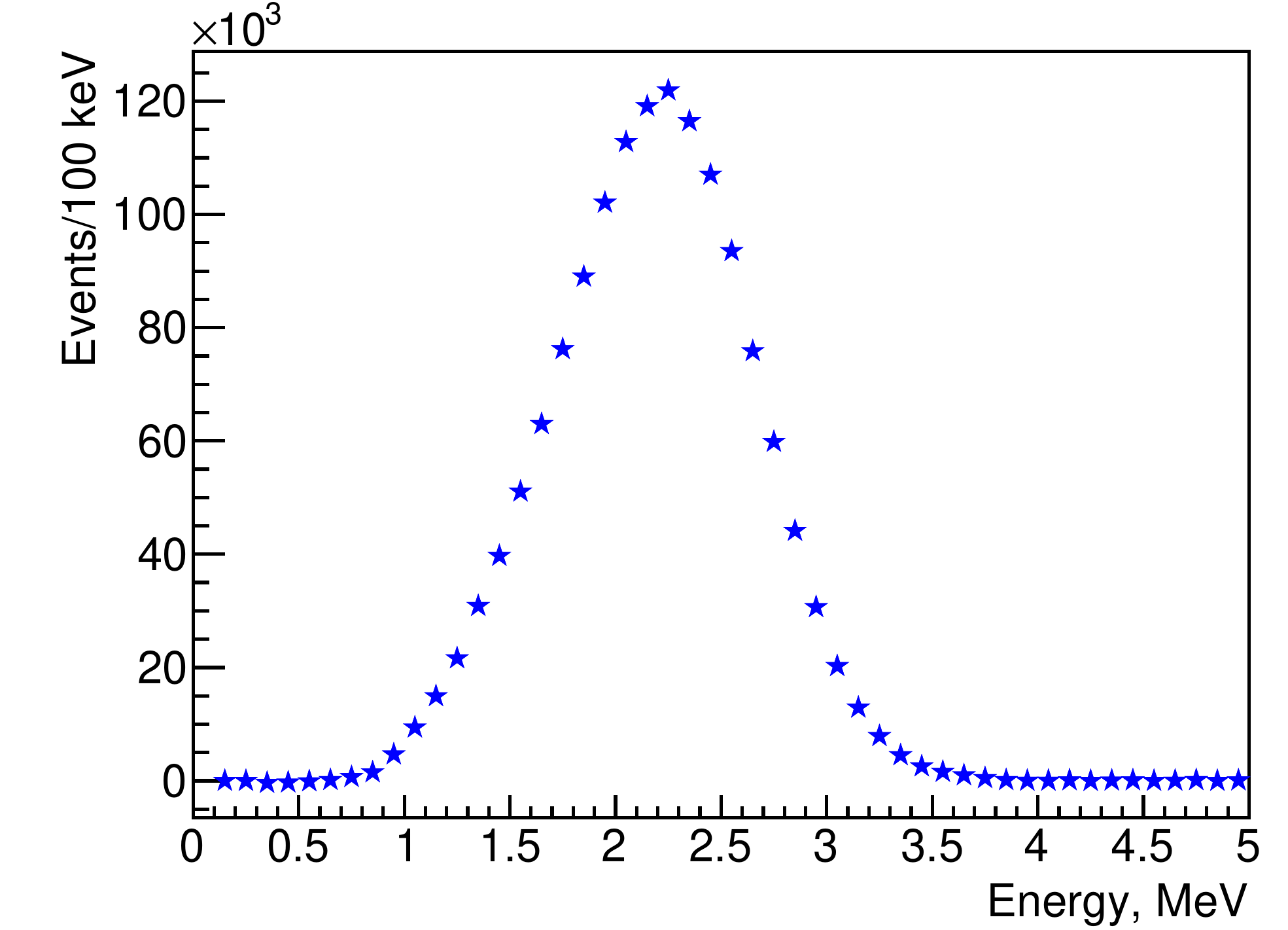}
\caption{\footnotesize Energy spectrum of the delayed signals measured with the $^{248}$Cm neutron source (left) and detector response to the $^{60}$Co source (right).}
 \label{Cm}
\end{figure}

Fig.~\ref{Cm}~(left) shows the energy distribution of neutron capture signals from a $^{248}$Cm source placed at the center of the detector. Two peaks correspond to the neutron capture by Gd and by protons. Fits of the high energy parts of the peaks give a resolution compatible with the MC simulations.
Fig.~\ref{Cm}~(right) shows the energy distribution of signals from a $^{60}$Co $\gamma$-source placed in the center of the detector. The observed energy resolution of $\frac {\sigma}{E}=20.3\%$ is consistent with the MC expectations.
The same is true in case of a $^{22}$Na source.

The SiPM gain calibration was performed using noise signals typically every 5 days. Calibration with cosmic muons of all strips in the whole detector was also performed once in $\sim 5$ days. A detailed description of the calibration procedure is presented elsewhere \cite{Calibration}. 
The strip response to the energy deposited by cosmic muons is linear within 0.7$\%$ in the range (1.7---4.7)~MeV. The energy measured by PMTs is proportional on average to the energy, measured by SiPMs. Therefore, the PMT energy response is also linear. Positrons with energies higher than 4.7~MeV typically deposit their energy in several strips. Therefore, the detector response should be linear for high energies as well.

\section{Data taking and analysis}
\begin{figure}

\includegraphics[width=0.45\textwidth]{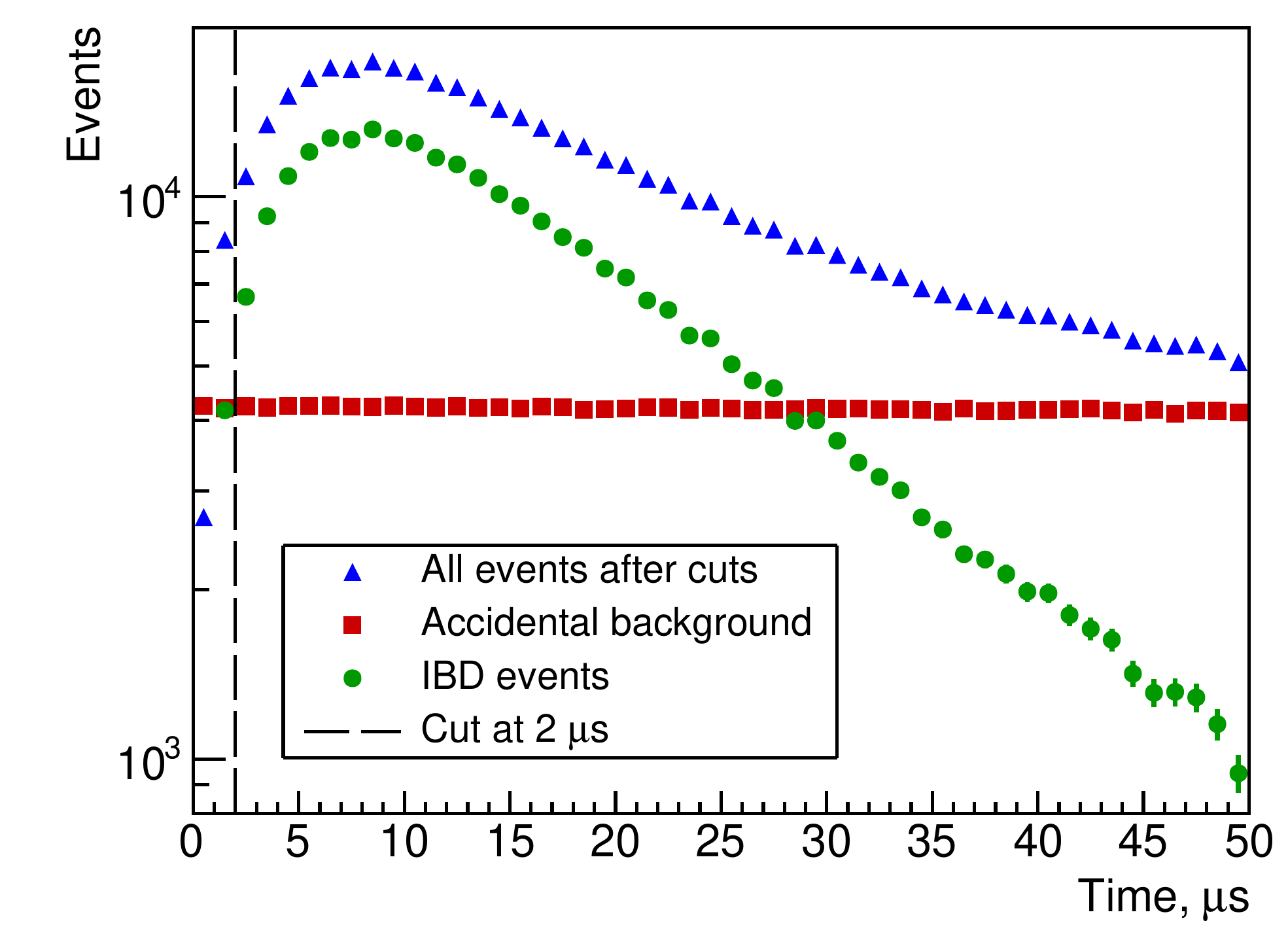}
\includegraphics[width=0.45\textwidth]{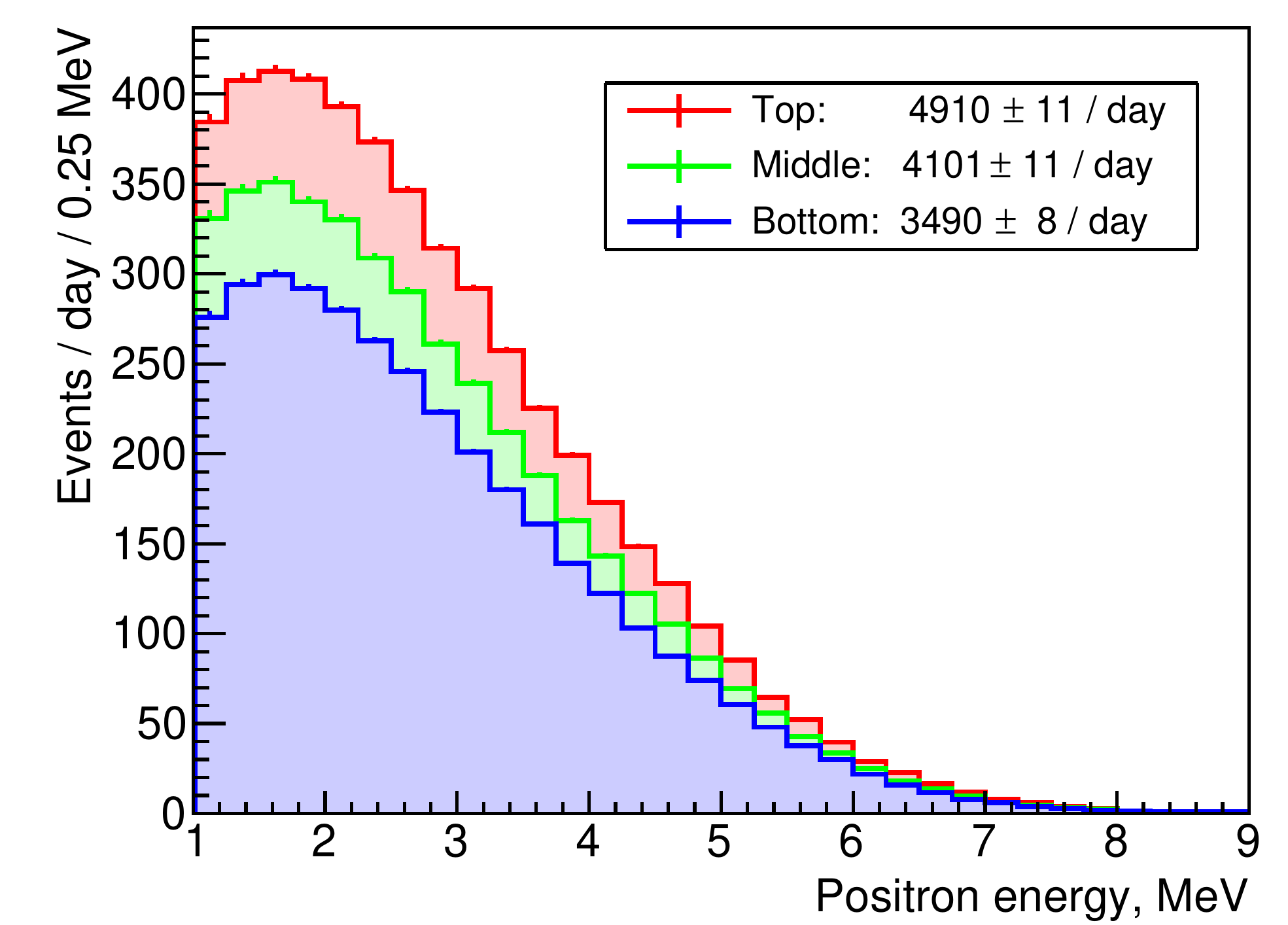}
\caption{\label{Time}Time between prompt and delayed signals (left) and positron energy distributions measured at different detector positions (right); 
statistical errors only.}
\end{figure}

The trigger of the experiment is produced when the digital sum of all PMT signals is above 0.7~MeV or the energy in the veto system is larger than 4~MeV. As a consequence, the IBD process appears in the data as two distinct events, prompt and delayed.  For each trigger, waveforms for all SiPMs and PMTs are recorded in 512~ns windows. 
 The visible energy of a positron cluster (continuous cluster of hits in the strips) is converted using MC simulations into the deposited energy by taking into account average losses in the inactive reflective layers of the strips and dead channels. Sometimes photons from the positron annihilation produce signals in the strips attributed to the positron cluster. This leads to an increase of the visible energy. Such a shift is also corrected on average using MC simulations.
A typical size of the total correction is $\sim2\%$.   
The next step is a search for the time-correlated pairs of prompt-delayed events.
We start with searching for an event with more than 3.5~MeV energy deposit. This is a delayed event candidate unless it has the muon veto. 
Then we look backward in time searching for a prompt event with more than 1~MeV in the positron cluster and no muon veto.
An IBD candidate pair is considered found if the time difference between the prompt and delayed events is in the range (2---50)~$\mu$s.
For a valid pair we also require no event with the muon veto within 60~$\mu$s before the prompt signal (within 200~$\mu$s if E~$>$~300~MeV is released in the main detector).  No other event should occur within 45~$\mu$s before
and 80~$\mu$s after the prompt event.
The found pairs of prompt and delayed events form the experimental sample of IBD candidates.
Similar to the experimental sample, the accidental coincidence sample is formed by looking for
a prompt signal in 16 regions: 5, 10,..., 80 ms before the neutron candidate. This sample provides us with a model-independent measure of the accidental background. 
Distributions for IBD candidates, the accidental background and their difference, which represents the IBD signal without accidental background are presented in fig.~\ref{Time}~(left).

Several cuts are applied in order to reduce the accidental background, for example a cut on the distance between positron and neutron candidates. These cuts are designed to be very soft with respect to the signal in order to avoid any distortions. 
Only the fiducial volume cut has a considerable inefficiency. The positron candidates are required to be at least 4 ~cm away from all detector edges.
All cuts were selected without looking at the final results. They have been fixed after collection of about 10$\%$ of the data. 

Positron energy spectra for 3 detector positions (top, middle, bottom) are shown in Fig.~\ref{Time}~(right)with
statistical errors only. The corresponding numbers of events are 338244, 247753, and 324938.
The IBD counting rate is 4910 events per day in the top position.
The positron energy does not include annihilation photons and hence is 1.02~MeV lower than the usually used prompt energy.  
The muon-induced background (discussed below) is subtracted.  
Three other reactors at the KNPP are 160~m, 334~m, and 478~m away from the DANSS detector.
The IBD counting rate from these reactors is 0.6$\%$ of the IBD counting rate from the nearest reactor for the top detector position. This contribution is taken into account by the corresponding reduction of the normalization of the obtained spectra.

The energy spectrum of the background from neutrons produced by muons inside the veto system is obtained from events with the muon veto. The amount of this background is determined from a fit of the positron candidate energy spectrum during reactor off periods using the shape of the background determined from events with the muon veto. This procedure reduces uncertainties in the background shape to a negligible level. A possible small uncertainty in the background rate is taken into account during systematic error studies. This is the most important background. It constitutes 2.7$\%$ of the IBD rate at the top detector position.

The shape of the positron spectrum agrees roughly with the MC predictions based on the $\anti_particle\nu_e$ spectrum from \cite{Huber, Mueller}. However, a quantitative comparison requires additional studies of calibration and systematic errors and improvements in the MC simulation of the detector. 

The shapes of the $\anti_particle\nu_e$ spectra are different for $^{235}$U and $^{239}$Pu isotopes. The fractions of these isotopes change during a reactor campaign. The amount of $^{235}$U decreases while the amount of $^{239}$Pu increases with time. This leads to changes in the $\anti_particle\nu_e$ spectrum first observed in\cite{Mikaelyan}. 

\begin{figure}[h]
\begin{center}
\includegraphics[width=0.55\textwidth]{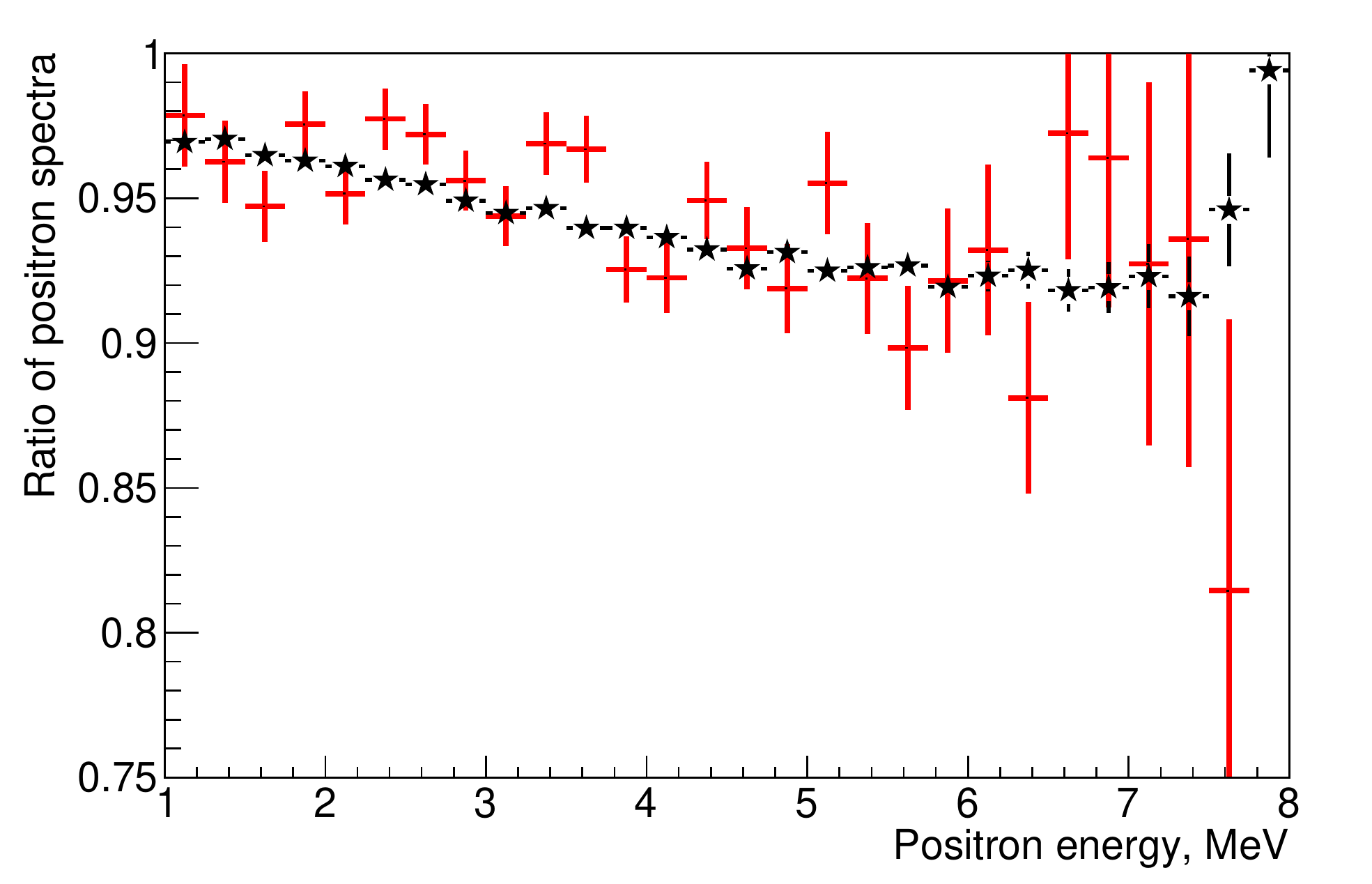}
\end{center}
\caption{\label{Evolution} Ratio of positron energy spectra collected during the last 3 months of the reactor campaign in 2017 and during the 2nd---4th months from the beginning of the next campaign (statistical errors only). The average $^{239}$Pu fission fractions for these periods are 37.7\% and 27.1\% correspondingly. Stars are the Hueber-Mueller model MC predictions. 
}
\end{figure}

Fig.~\ref{Evolution} shows the ratio of positron spectra collected during 3  months before the end of the reactor campaign and during the 2---4 months after beginning of the campaign. The corresponding fractions of $^{239}$Pu ssion fractions are 37.7\% and 27.1\%
correspondingly.
We do not have reliable information about the corresponding fraction during the first month after the start of the reactor campaign. The changes in the positron spectrum are obvious. They are well described by the MC expectations based on the $\anti_particle\nu_e$ spectra from \cite{Huber, Mueller} contrary to the Daya Bay measurements \cite{DBEvolution} which exhibit smaller changes than the MC predictions. 

\begin{figure}[h]
\begin{center}
\includegraphics[width=0.95\textwidth]{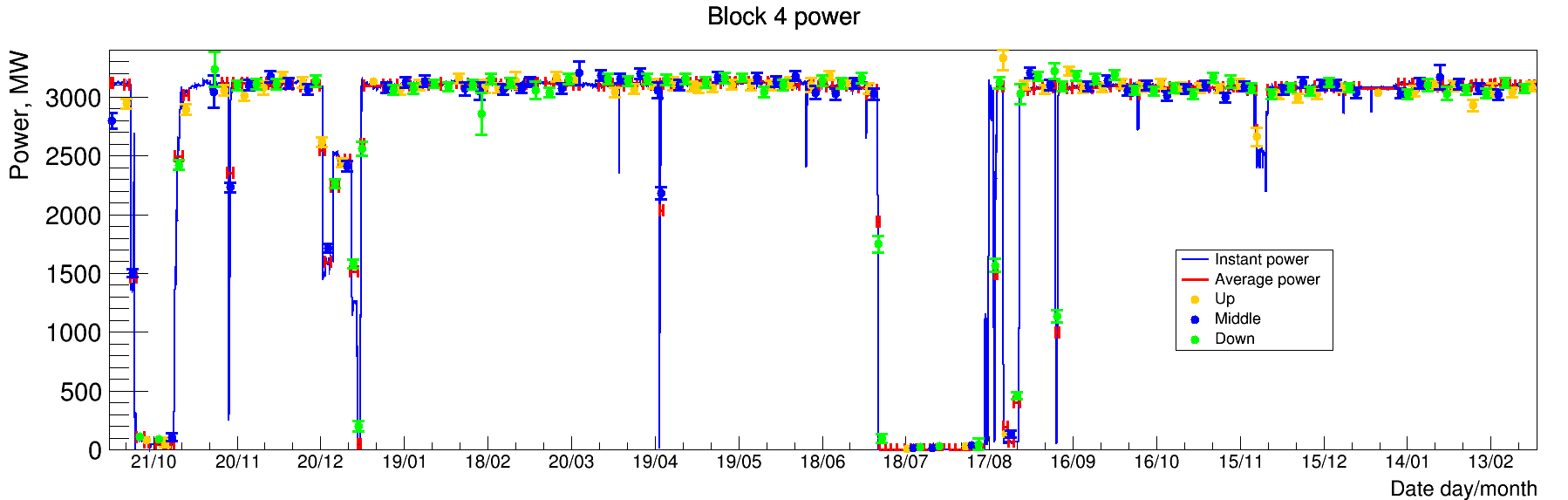}
\end{center}
\caption{\label{Power} IBD counting rate at different detector positions (points) and the reactor thermal power. The distributions are equalized during the period November 15 – December 16, 2016 and therefore the IBD rate is expressed in MW. The IBD rate is corrected for the fuel evolution with time using Huber-Mueller model. Statistical errors only.}
\end{figure}

Fig.~\ref{Power} shows the correlation between the reactor thermal power and the number of registered IBD events. The two distributions were equalized during one month period in 2016. The number of IBD events was corrected for the fuel evolution using the model of the  $\anti_particle\nu_e$ spectra \cite{Huber, Mueller}. After that the IBD rate coincides with the reactor power with about 2$\%$ accuracy during 16 months. The only exception is the one month period after the latest beginning of the campaign for which we do not have reliable data on the isotope fractions.

\begin{figure}[h]
\begin{center}
\includegraphics[width=0.55\textwidth]{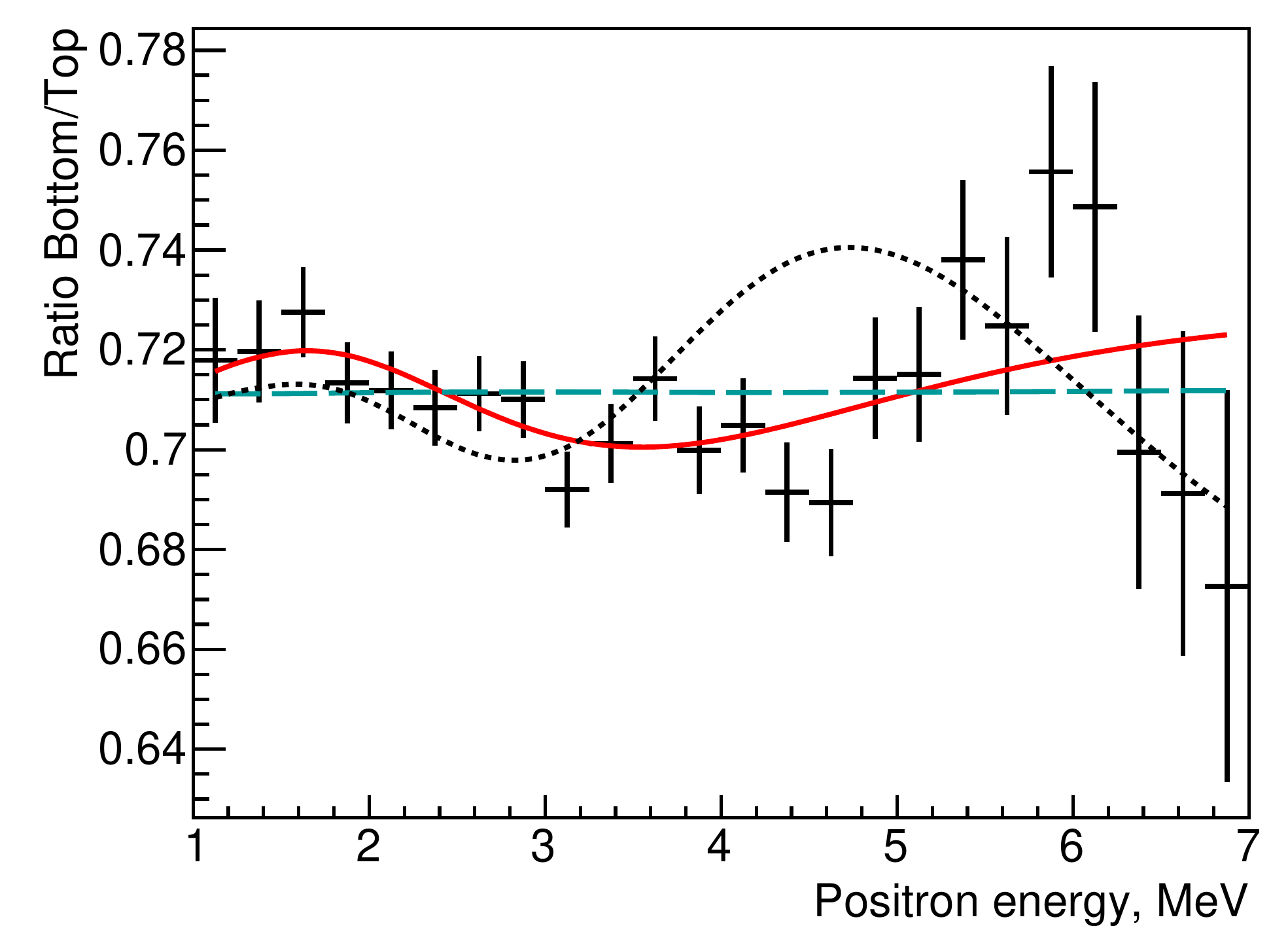}
\end{center}
\caption{\label{Ratio}Ratio of positron energy spectra measured at the bottom and top detector positions (statistical errors only). The dashed line is the prediction for 3$\nu$ case ($\chi^2=35.0$, 24 degrees of freedom). The solid curve is the prediction for the best fit in the $4\nu$ mixing scenario ($\chi^2 = 21.9$,~$ \sin^22\theta_{14} = 0.05$,~$\Delta m_{14}^2 = 1.4~\rm{eV}^2$). The dotted curve is the expectation for the optimum point from the RAA and GA fit \cite{Mention2011} ($\chi^2=83$, $\sin^22\theta_{14}=0.14$,~$\Delta m_{14}^2 = 2.3~\rm{eV}^2$)}
\end{figure}

Fig.~\ref{Ratio} shows the ratio of positron energy spectra at the bottom and top detector positions. 
The exclusion area in the sterile neutrino parameter space was calculated using the Gaussian \cls method \cite{CLS} assuming only one type of sterile neutrinos. For a grid of points in the $\Delta m_{14}^2, \sin^22\theta_{14}$ plane predictions for the ratio $R^{pre}(E)$  of positron spectra at the bottom and top detector positions were calculated. Calculations included the MC integration over the $\anti_particle\nu_e$  production point in the reactor core,  $\anti_particle\nu_e$  detection point in the detector, and positron energy resolution. The $\anti_particle\nu_e$  production point distributions in the reactor core were provided by the KNPP for different time periods.   The distribution averaged over the campaign was used in the calculations. It was checked that this approximation practically did not influence the final results. 

The obtained theoretical prediction for a given point in the $\Delta m_{14}^2,~\sin^22\theta_{14}$ plane was compared with the prediction for the three neutrino case using the Gaussian \cls method for the $90\%$ confidence level (CL) exclusion area estimation. The difference in $\chi^2$ for the two hypotheses $\Delta\chi^2 = \chi^2_{4\nu} - \chi^2_{3\nu}$ was used for the comparison.  The \cls method is more conservative than the usual Confidence Interval method. 

The $\chi^2$ for each hypothesis was constructed using 24 data points $R^{obs}_i$ in the (1---7)~MeV positron energy range 

\begin{equation}
\label{eq2}
\chi^2 = \sum_{i=1}^N(R^{obs}_i-k \times R_i^{pre})^2/\sigma_i^2,
\end{equation}
where $R^{obs}_i$ ($R^{pre}_i$) is the observed (predicted) ratio of $\anti_particle\nu_e$ counting rates at the two detector positions and $\sigma_i$ is the statistical standard deviation of $R^{obs}_i$, and $k$ is a normalization factor left free in the fit to compensate possible changes in the overall detector efficiency and in the $\anti_particle\nu_e$  production point distribution in the reactor core.   

 The oscillations due to the known neutrinos were neglected since at such short distances they do not change the $\anti_particle\nu_e$ spectrum. 
 The procedure was repeated for all points of the grid in order to get the whole exclusion area. 
Influence of systematic uncertainties in the parameters was estimated by repeating the analysis with different values of parameters. A point in the $\Delta m_{14}^2,~\sin^22\theta_{14}$ plane was included into the final excluded area if it appeared in the excluded areas for all tested variations of the parameters. We varied the energy resolution and the background level by 10$\%$ and 15$\%$ correspondingly. We also used a reduced energy range of (1.5---6)~MeV in the fit.

\begin{figure}[h]
\begin{center}
\includegraphics[width=0.6\textwidth]{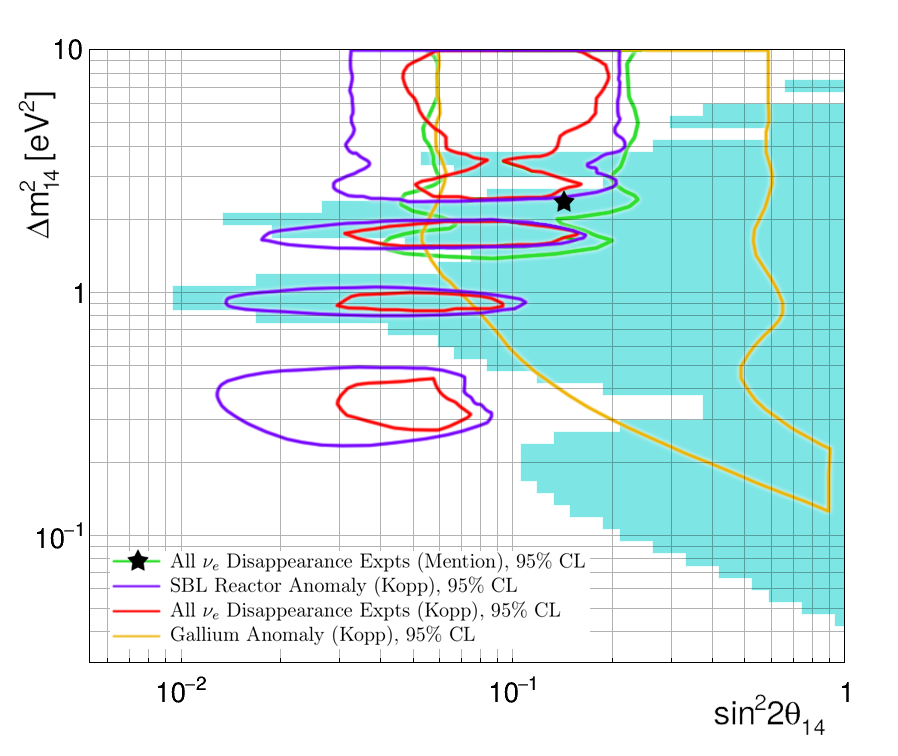}
\end{center}
\caption{\label{Exclusion}90$\%$ CL exclusion area in $\Delta m_{14}^2,~\sin^22\theta_{14}$ parameter space. The shaded area represents our analysis. Curves show allowed regions from neutrino disappearance experiments\cite{Mention2011,contours}, and the star is the best point from the RAA and GA fit\cite{Mention2011}.}
\end{figure}

Fig.~\ref{Exclusion} shows the obtained 90$\%$ CL excluded area in the $\Delta m_{14}^2,~\sin^22\theta_{14}$ plane. 
For some values of $\Delta m^2_{14}$ the obtained limits are more stringent than previous results\cite{Bugey,DBoscillations,NEOS}. It is important to stress that our results are based only on the comparison of the positron energy distributions at the two distances measured with the same detector.
Therefore the results do not depend on the $\anti_particle\nu_e$ spectrum shape and normalization as well as on the detector efficiency.  The excluded area covers a large fraction of regions indicated by the GA and RAA. In particular, the most preferred point $\Delta m_{14}^2=2.3~\rm{eV}^2,~\sin^22\theta_{14} =0.14$ \cite{Mention2011} is excluded at more than 5~$\sigma$ CL. 
In our analysis the point $\Delta m_{14}^2=1.4~\rm{eV}^2,~\sin^22\theta_{14} =0.05$ has the smallest $\chi^2 = 21.9$. The difference in $\chi^2$ with the 3$\nu$ case is 13.1 which corresponds to $\sim3\sigma$ in case of the $\chi^2$ distribution with 2 degrees of freedom. The significance of this indication of the existence of the sterile neutrino will be studied taking into account systematic uncertainties after collection of more data this year.

\section*{Acknowledgments}
We appreciate the permanent assistance of the KNPP administration and Radiation and Nuclear Safety Departments. The detector construction was supported by the Russian State Corporation ROSATOM (state contracts H.4x.44.90.13.1119 and H.4x.44.9B.16.1006). The operation and data analysis became possible due to the valuable support from the Russian Science Foundation grant 17-12-01145. The preparation of this paper was supported by the Russian Federal Government grant 14.W0331.0026.

%
%
%

\end{document}